# Observing the Birth of Rydberg Exciton Fermi Polarons on a Moiré Fermi Sea

Eric A. Arsenault[1,*], Gillian E. Minarik[1], Jiaqi Cai[2], Minhao He[2], Yiliu Li[1], Takashi Taniguchi[3], Kenji Watanabe[4], Dmitri Basov[5], Matthew Yankowitz[2,6], Xiaodong Xu[2,6], and X.-Y. Zhu[1,*]

[1] Department of Chemistry, Columbia University, New York, NY 10027, USA
[2] Department of Physics, University of Washington, Seattle, WA 98195, USA
[3] Research Center for Materials Nanoarchitectonics, National Institute for Materials Science, 1-1 Namiki, Tsukuba 305-0044, Japan
[4] Research Center for Electronic and Optical Materials, National Institute for Materials Science, 1-1 Namiki, Tsukuba 305-0044, Japan
[5] Department of Physics and Astronomy, Columbia University, New York, NY 10027, USA
[6] Department of Materials Science and Engineering, University of Washington, Seattle, WA 98195, USA

**The optical spectra of two-dimensional (2D) semiconductors are dominated by tightly bound excitons and trions. In the low doping limit, trions are often described as three-body quasiparticles consisting of two electrons and one hole or vice versa. However, trions are more rigorously understood as quasiparticles arising from the interaction between an exciton and excitations of the Fermi sea—referred to as exciton Fermi polarons. Here, we employ pump-probe spectroscopy to directly observe the formation of exciton Fermi polarons in a model system composed of a $WSe_2$ monolayer adjacent to twisted bilayer graphene (tBLG). Following the pump injection of Rydberg excitons in $WSe_2$, a time-delayed probe pulse tracks the development of Rydberg exciton Fermi polarons as interactions with localized carriers in the tBLG moiré superlattice evolve. Both the exciton Fermi polaron relaxation rate and the binding energy are found to increase with electron or hole density. Our findings provide insight into the optical response of fundamental excitations in 2D van der Waals systems and reveal how many-body interactions give rise to emergent quasiparticles.**

---

[*] To whom correspondence should be addressed. EAA: eaa2181@columbia.edu; XYZ: xyzhu@columbia.edu.

**Introduction**

Arising from reduced dimensionality, screening effects in two-dimensional (2D) semiconducting transition metal dichalcogenides (TMDs) are significantly diminished [1–4]. This reduced screening enhances Coulomb interactions between charge carriers, leading to an increase in both the quasiparticle band gap and the binding energy of excitons (i.e., bound electron-hole pairs), as well as the emergence of many-body phenomena. Consequently, the optical spectra of 2D TMD monolayers are dominated by tightly bound excitons [5,6] and so-called trions [7–9]. While a trion is often described as a three-body quasiparticle consisting of either two electrons and one hole or the opposite charge configuration [2], this approximation is appropriate only at low doping densities [10]. Rather, the corresponding quasiparticle is more rigorously considered as an exciton dressed by excitations (i.e., density fluctuations) of the Fermi sea [11,12], forming what is known as an exciton Fermi polaron [11–17]. Trions or exciton Fermi polarons have been observed not only to form from the 1s exciton in doped monolayers, but also from Rydberg excitons in TMD monolayers [17–19], as well as from interlayer and moiré-trapped excitons in TMD bilayers [20–23]. While these quasiparticles have been probed in static optical measurements [7,8,12], experiments offering dynamic views are limited [15,16], mostly to the transition or coupling within the two-state exciton-trion picture [9,24–29]. However, the exciton Fermi polaron picture inherently involves dynamical behavior. For example, the response of the Fermi sea to the presence of an exciton, as well as the relaxation of energy within the electronic system, occur on characteristic timescales determined by underlying many-body interactions [30–34]. This many-body-dependent dynamic response, which is also essential to the understanding of other polaron problems [34–36], is our focus here.

To directly capture the birth of exciton Fermi polarons, we choose the model system of a monolayer $WSe_2$ interfaced with twisted bilayer graphene (tBLG) [37,38]. Here, the excitonic degrees of freedom reside in $WSe_2$, atop a gate-tunable moiré Fermi sea contained within tBLG. As the Fermi level lies deep in the bandgap of $WSe_2$, electrostatic doping selectively tunes the charge density in the tBLG moiré potential. Importantly, the presence of the moiré potential enhances both the achievable quasi-local charge density and the strength of interlayer Coulomb interactions between the excitons and periodic charge density [37,38]. As a result, moiré Rydberg excitons with principal quantum numbers $n \geq 2$ and enhanced oscillator strength have been observed in this system, likely a consequence of symmetry breaking [37,38]. Altogether, this model system is particularly advantageous for observing the formation of exciton Fermi polarons. In an ideal Rydberg system [39], the exciton radius (and thus the dipole moment) scales as $n^2$ while the polarizability scales as $n^7$. Although deviations from this ideal behavior exist for 2D TMDs [40], the $n^2$ scaling approximately holds [19]. Compared to the commonly observed 1s (i.e., $n = 1$) exciton state, Rydberg excitons possess larger dipole moments and polarizabilities, affording stronger Coulomb interactions with the Fermi sea. This leads to larger polaron binding energies and facilitates more straightforward experimental observation. To establish the dynamics of Rydberg exciton Fermi polaron formation, we apply pump-probe spectroscopy. In this approach, a pump pulse excites Rydberg excitons in the $WSe_2$ monolayer; following a controlled time delay, a probe pulse tracks the energy and amplitude evolution of the Rydberg Fermi polaron state as it interacts with carriers in the tBLG moiré superlattice. By tuning the carrier density via doping of the tBLG, we demonstrate the strong dependence of Rydberg Fermi polaron formation dynamics on electron or hole density, providing insight into how many-body interactions govern the emergence of quasiparticle states in 2D devices.

**Results and Discussion**

A schematic of the $WSe_2$/tBLG device is shown in Fig. 1a. The sample features a dual-gated Hall bar geometry to allow for control over the carrier density, $\rho$, as well as the implementation of transport measurements. Each gate consists of a hexagonal boron nitride (h-BN) dielectric spacer and few-layer graphene (fl-Gr) electrode. Initial device characterization consisted of transport measurements to determine the tBLG twist angle ($\theta_{tBLG} = 0.8°$) and to calibrate the gate-dependent $\rho$. By convention, positive (negative) $\rho$ indicates electron (hole) doping. At this twist angle, the moiré landscape provides a periodic potential for doped charge, without the formation of correlated insulator states [41–44] that would otherwise complicate the net $\rho$-dependent influence on exciton resonances in the adjacent $WSe_2$ monolayer [37,38]. Additionally, Hu et al. demonstrated that at $\theta_{tBLG} = 0.8°$, there is a near-optimal balance between strong interlayer coupling and preservation the $WSe_2$ 2s oscillator strength [37]. Further details on sample fabrication and transport measurements can be found in Methods and Fig. A1.

The tBLG moiré superlattice provides a gate-tunable, spatially periodic charge density that imprints on the neighboring $WSe_2$ monolayer. The ramifications of this adjacent superlattice can be tracked directly via the $\rho$-dependent optical response, Fig. 1b. As detailed previously [37,38], disparate behavior is observed in the WSe$_2$ 1s versus 2s exciton regimes (1.69~1.71 eV and 1.74~1.79 eV, respectively, Fig. 1b). While the 1s exciton resonance is $\rho$-independent, the WSe$_2$ 2s exciton resonance undergoes a major spectral redshift, as well as peak splitting, as $|\rho|$ increases. The absence of an energy shift for the WSe$_2$ 1s exciton arises from a cancellation between the screening-induced reduction in exciton binding energy and the renormalization of the quasiparticle bandgap [45–47]. In contrast, for the 2s Rydberg exciton, the reduction in binding energy is insufficient to fully compensate for bandgap renormalization, leading to a net redshift. Additionally, compared to the 1s exciton, the increased dipole moment and polarizability of Rydberg excitons give rise to much stronger Coulomb interactions between excitons in WSe$_2$ and charge carriers in tBLG, allowing us to track the dynamic response of the moiré Fermi sea during the formation of Rydberg exciton polarons, as illustrated in Fig. 1c.

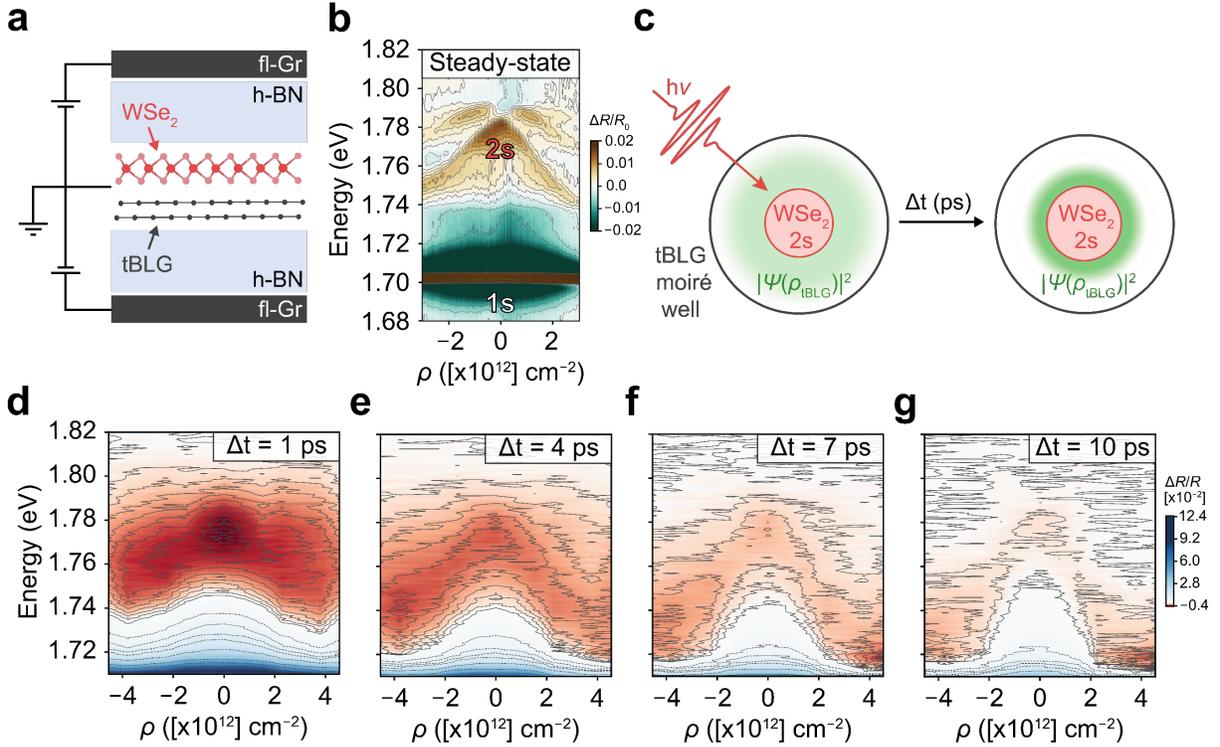

**Figure 1. WSe$_2$/tBLG Device Architecture and Characterization of Moiré Rydberg Excitons.**
**a** Schematic of device architecture where the WSe$_2$/tBLG stack is symmetrically encapsulated by top and bottom gates consisting of h-BN and fl-Gr. **b** Steady-state reflectance spectrum as a function of carrier density, $\rho$, at T = 5 K. Here, $\Delta R = R - R_0$ where $R$ is the reflected signal from the active region while $R_0$ is the reflected signal from that same region under high $\rho$ as there was otherwise no suitable background region in the sample. **c** Cartoon of exciton Fermi polaron formation. Here, a pump pulse injects a Rydberg exciton (red circle) in the WSe$_2$ layer suspended above the Fermi sea (charge density shown in green) contained in the tBLG moiré potential well (simplistically represented by the gray circle). Dynamic many-body interactions between the photoinjected Rydberg exciton and moiré Fermi sea lead to exciton Fermi polaron formation, as captured by the enhanced localization of the charge density (shown in green) during the waiting time, $\Delta t$. **d-g** Time-resolved reflectance response plotted as a function of probe energy (covering the moiré Rydberg exciton region) versus $\rho$ at specific pump-probe delays ($\Delta t$ = 1, 4, 7, and 10 ps in panels d-g, respectively). Here, $\Delta R = R_{on} - R_{off}$ where the subscript indicates either pump-on or -off spectra. Data collected at T = 14 K. See Methods for further details. Pump fluence is 38 μJ/cm$^2$ for all transient data.

Figs. 1d-g display snapshots of reflectance versus $\rho$ at selected pump-probe delays ($\Delta t$ = 1, 4, 7, 10 ps). In time-resolved reflectance measurements, an optical pump pulse with an above gap photon energy of $h\nu_1$ = 1.91~2.07 eV populates excitons in WSe$_2$. The probe pulse spanning $h\nu_2$ = 1.71~1.82 eV tracks the Rydberg exciton dynamics. In the time-resolved reflectance maps, a photobleach feature due to Pauli blocking appears for the lowest moiré Rydberg exciton branch of mainly 2s character. Other moiré Rydberg states at energies above that of the dominant branch, as observed in the steady-state measurement in Fig. 1b, are not resolved in the pump-probe measurements (Fig. 1d-g). This is likely a result of excitation-induced broadening [48,49]. There are two notable features in the coarse temporal mapping of the 2s Rydberg exciton with increasing $|\rho|$: i) the peak position undergoes a more significant redshift with $\Delta t$ and ii) the recovery dynamics appear to become slower. In the following, we address the mechanistic origin of these features. See Methods for a discussion of the unavoidable photoexcitation of tBLG, with dynamics occurring on a sub-2 ps timescale, significantly faster than the two aforementioned notable features.

To carefully assess these dynamics, we plot the energy-resolved probe response of the lowest energy moiré Rydberg state as a function of $\Delta t$ under increasing hole (Fig. 2) and electron (Fig. 3) doping. Overlaid on each pseudocolor plot is the spectral evolution of the 2s moiré Rydberg exciton transition energy tracked as a function of $\Delta t$. At a given $\Delta t$, the exciton energy is taken to be the energy corresponding to the maximum peak position (given by the circles in Fig. 2 and 3) where the black vertical lines denote the energy range that lies within 95% of the peak maximum. Further details are provided in Methods and Fig. A2-A4. At charge neutrality ($\rho$ = 0 cm$^{-2}$), the excitonic peak position notably remains constant with increasing $\Delta t$. Similar behavior is observed under moderate hole or electron doping ($|\rho|$ = 0.8x10$^{12}$ cm$^{-2}$). However, at higher doping levels ($|\rho| \geq$ 1.5 x10$^{12}$ cm$^{-2}$), the exciton energy is observed to undergo a transient redshift—the magnitude of which increases with doping. At the highest hole doping levels ($\rho$ = -3.8x10$^{12}$ cm$^{-2}$ and -4.5 x10$^{12}$ cm$^{-2}$), the dynamic peak redshift is on the order of ~30 meV at $\Delta t \geq$ 10 ps. For the same magnitude of electron doping, the dynamic redshift is sufficiently large (~35 meV at $\Delta t \geq 10$ ps) that the 2s exciton branch seems to merge into tail of the intense 1s exciton transition. Clearly, the three-body interpretation of the trion picture is inadequate to account for the observed doping-dependent behavior. Rather, the time-dependent redshift of the Rydberg exciton energy and its strong dependence on charge density in the tBLG moiré lattice provides a direct view of polaron formation, i.e., the relaxation of the surrounding environment due to the presence of a quasiparticle [30–36]. In the present case, the quasiparticle is the Rydberg exciton in $WSe_2$ and the environment is the moiré Fermi sea in the adjacent tBLG, Fig. 1c. While this cartoon illustrates an interaction between a Rydberg exciton and the charge density at a single moiré site, this representation is a simplification and may not strictly reflect the physical scenario. Given that the charge density is not fully localized, polaron formation may reasonably involve contributions from neighboring sites. The spatial extent over which the Fermi sea participates in this interaction remains an open question that warrants further investigation. We also note that this transient redshift is in addition to the $\rho$-dependent static redshift of the moiré Rydberg exciton transitions already captured in the steady-state spectrum (and the early time transient data), which is primarily attributed to dielectric screening effects [37,38].

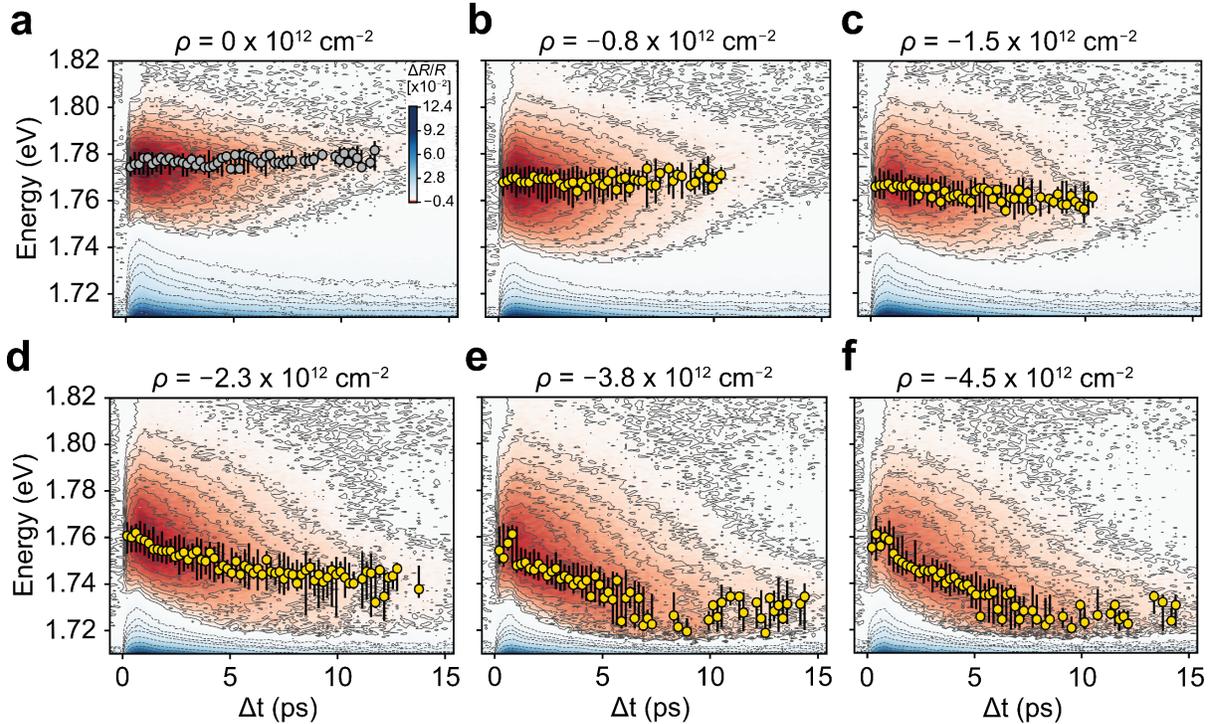

**Figure 2. Transient Response of the Moiré Rydberg Exciton State as a Function of Hole Doping.**
**a-f** Carrier density ($\rho$)-dependent exciton response shown as a function of probe energy versus pump-probe delay ($\Delta t$). Here, $\Delta R = R_{on} - R_{off}$ where the subscript indicates either pump-on or -off spectra. In each panel, the $\Delta t$-dependent peak shift is shown on top of the corresponding pseudocolor plot. At a given $\Delta t$, the moiré Rydberg exciton transition energy is taken to be the energy at the maximum peak position. The black vertical lines denote the energy range that lies within 95% of the determined peak maximum. In panel a, the extracted peak frequencies are shown in gray circles to distinguish the charge neutral data, while the yellow circles in panels b-f distinguish the transient peak shifts under hole doping. We note that under high hole doping, the significant peak redshift toward the edge of the probe window prohibited the extraction of a reliable center frequency at all $\Delta t$. As a result, such points were omitted. See Methods for further details. All data collected at T = 14 K.

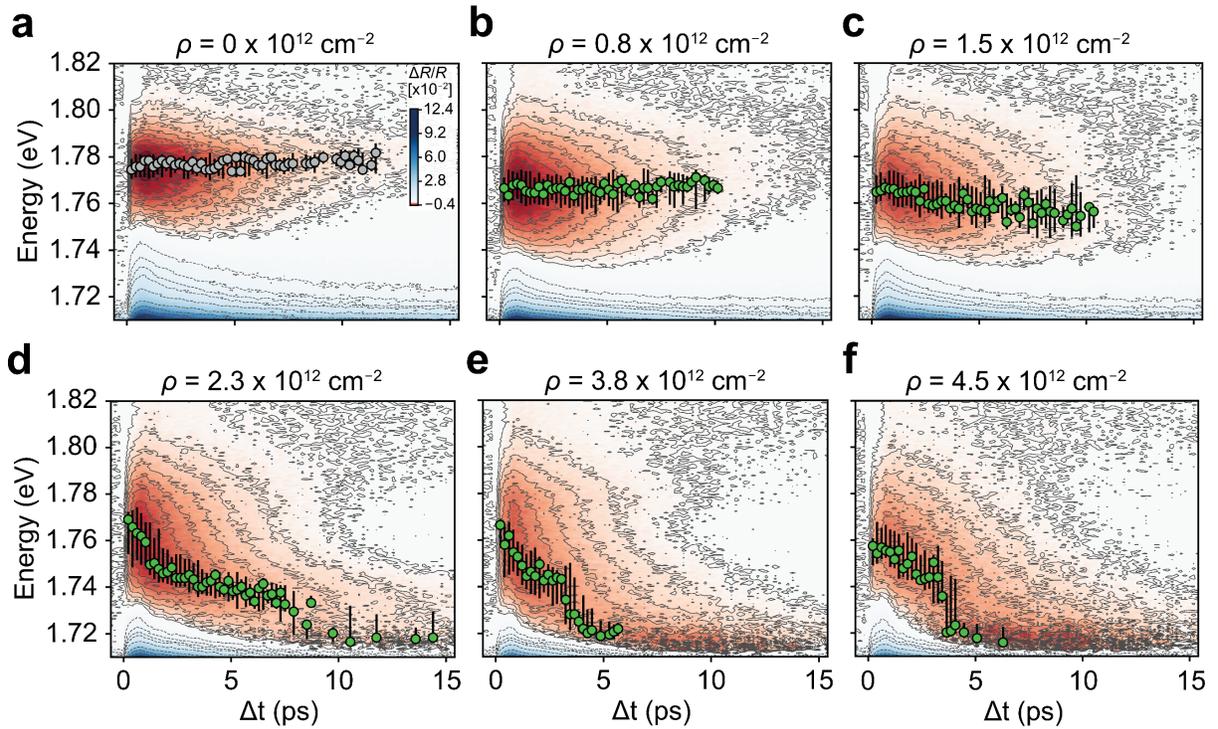

**Figure 3. Transient Response of the Moiré Rydberg Exciton State as a Function of Electron Doping. a-f** Carrier density ($\rho$)-dependent exciton response shown as a function of probe energy versus pump-probe delay ($\Delta t$). Here, $\Delta R = R_{on} - R_{off}$ where the subscript indicates either pump-on or -off spectra. In each panel, the $\Delta t$-dependent peak shift is shown on top of the corresponding pseudocolor plot. At a given $\Delta t$, the moiré Rydberg exciton transition energy is taken to be the energy at the maximum peak position. The black vertical lines denote the energy range that lies within 95% of the determined peak maximum. In panel a, the extracted peak frequencies are shown in gray circles to distinguish the charge neutral data, while the green circles in panels b-f distinguish the transient peak shifts under electron doping. Panel a, identical to Fig. 2a, is repeated here to highlight the observed $\rho$-dependent behavior. We note that under high electron doping, the significant peak redshift toward the edge of the probe window prohibited the extraction of a reliable center frequency at all $\Delta t$. As a result, such points were omitted. See Methods for further details. All data collected at T = 14 K.

We now consider two important results: the overall redshift observed in the long-time limit and the rate of energy relaxation. The former provides a measure of the Fermi polaron binding energy, which is ~30 meV and ≥ 35 meV (where this is a lower bound due to spectral overlap near the edge of the probe window) for the highest measured hole and the electron carrier densities, respectively. For the latter, we extract the rate of energy relaxation, $k_{relaxation}$, for each $\rho$ (see Methods and Fig. A5-A6 for further details). The sign convention is such that a negative $k_{relaxation}$ indicates a redshift with respect to the static transition. As shown in Fig. 4a, the magnitude of energy relaxation rate, $|k_{relaxation}|$, is negligible for $|\rho| \leq 0.8 \times 10^{12}$ cm$^{-2}$ and increases with $|\rho|$ at higher doping levels. The energy relaxation rate reflects the collective response of the Fermi sea, or more succinctly the local Fermi "pond" (referring to the enhanced local charge density induced by the moiré potential), on the tBLG moiré landscape to the presence of the Rydberg exciton in the WSe$_2$ monolayer. This response likely corresponds to the timescale for screening, and thus to the 2D plasmon frequency of the electrons or holes, as indicated by the $\rho$-dependent energy relaxation rates. However, in tBLG, the behavior of 2D plasmons is modified in major ways due to band flattening and presence of the moiré superlattice potential [50,51], complicating quantitative comparisons in the present study. Interestingly, we note that $k_{\text{relaxation}}$ is not symmetric under electron versus hole doping. This asymmetry may arise from differences between the valence and conduction moiré bands, leading to distinct effective masses and degrees of carrier localization.

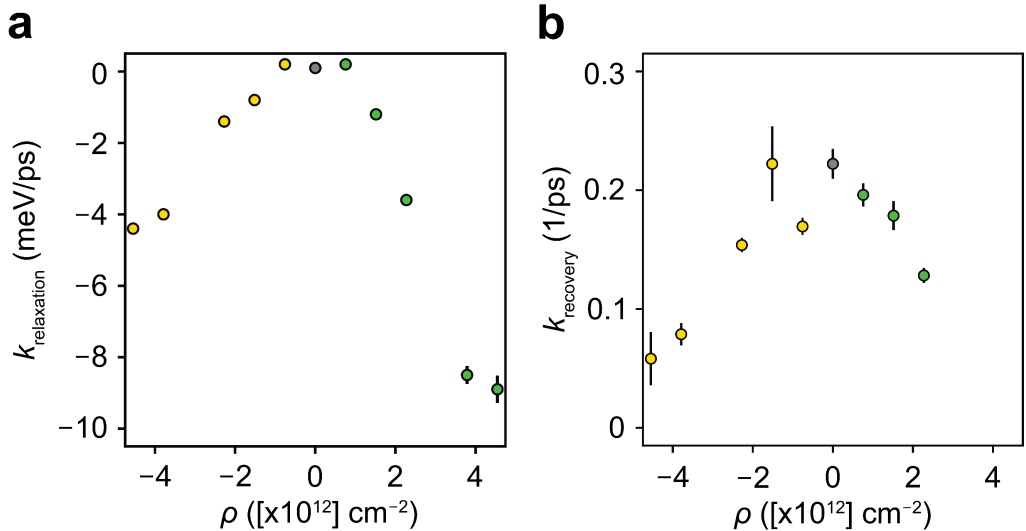

**Figure 4. Doping-Dependent Dynamics of Rydberg Exciton Fermi Polaron Formation. a** Rate of energy relaxation, $k_{relaxation}$, for the moiré Rydberg exciton transition as a function of carrier density, $\rho$. Rates were extracted via linear fits of the peak position as a function of energy. The corresponding data is shown in the shaded circles in Fig. 2 and 3. We note that the negative rate of energy relaxation indicates a redshift with respect to the static moiré Rydberg exciton transition energy. **b** Rate of recovery, $k_{recovery}$, for the moiré Rydberg exciton transition is obtained via fitting the amplitude profiles of the dynamic peak positions as a function of $\Delta t$ with an exponential recovery. $k_{recovery}$ for the two highest electron doping levels are omitted due to large uncertainties associated with extracting peak positions at the edge of the probe window. Throughout the figure, hole (electron) doping is indicated in yellow (green) and charge neutral data is shown in gray. Error bars indicate 99% confidence intervals. See Methods for further details and Fig. A5-A7 for corresponding fits.

Qualitatively, the magnitude of $k_{\text{relaxation}}$ can be taken as a proxy for the degree of interaction between a Rydberg exciton in WSe$_2$ and the Fermi "pond" on the tBLG moiré landscape. Stronger interactions during exciton Fermi polaron formation are expected to result in more effective screening of the Rydberg exciton, leading to a decreased exciton recombination rate. To confirm the formation of exciton Fermi polarons, Fig. 4b shows at each $\rho$ the rate of recovery, $k_{\text{recovery}}$, which is obtained from fitting each amplitude profile to an exponential decay (see Methods and Fig. A7 for further details). The slowing down of $k_{\text{recovery}}$ with increasing $|\rho|$ confirms the screening effect.

**Conclusion**

In summary, we transiently resolve the formation of Rydberg exciton Fermi polarons on the moiré landscape in a model system composed of a WSe$_2$ monolayer on top of tBLG. We observe that the lowest-energy moiré Rydberg exciton branch exhibits $\rho$-dependent transient redshifts, accompanied by characteristic rates of energy relaxation and recovery—key signatures of polaronic behavior. These features capture the dressing of Rydberg excitons by the Fermi "pond," resulting in dynamic renormalization of exciton energies driven by many-body correlations. This work not only establishes the carrier density regime over which the exciton Fermi polaron picture manifests in a moiré system, but also offers broader insight into how dynamic many-body interactions govern the formation of quasiparticle states. More broadly, the reduced dimensionality of van der Waals materials enhances Coulomb interactions, amplifying many-body effects that often dictate their optical response. Understanding how these interactions dynamically emerge is therefore essential for advancing the control and design of optoelectronic phenomena in the 2D limit.

**Acknowledgements**

The spectroscopic work was supported by Department of Defense Multidisciplinary University Research Initiative grant number W911NF2410292. Additional supports for sample fabrication came from Programmable Quantum Materials, an Energy Frontier Research Center funded by the U.S. Department of Energy (DOE), Office of Science, Basic Energy Sciences (BES) under award number DE-SC0019443, and for methodology/instrument development by DOE-BES under award DE-SC0024343. EAA gratefully acknowledges support from the Simons Foundation as a Junior Fellow in the Simons Society of Fellows (965526). Facilities supported by the Materials Science and Engineering Research Center (MRSEC) through NSF grant DMR-2011738 were utilized in this work. K.W. and T.T. acknowledge support from the JSPS KAKENHI (Grant Numbers 21H05233 and 23H02052) and World Premier International Research Center Initiative (WPI), MEXT, Japan. We thank Yinjie Guo for assistance with sample mounting.

**Competing Interests**

The authors declare no competing interests.

**Data Availability Statement**

The data that support the findings of this article are openly available [52].

## Appendix A: Methods

### Sample Fabrication and Calibration

The WSe$_2$/tBLG heterostructures are assembled using a standard dry-transfer technique with a polycarbonate (PC) stamp and the tear-and-stack method. First, graphite and h-BN are mechanically exfoliated onto a SiO$_2$/Si wafer. Graphite, graphene, WSe$_2$, and h-BN flakes with appropriate thickness are identified with an optical microscope and confirmed clean with an atomic force microscope or a color-contrast-enhanced optical microscope. We select 3-5 layers of graphite as the top gate to enhance optical transparency, and thicker 3-5 nm graphite as the bottom gate. Then, the PC stamp is used to pick up graphite, h-BN, and monolayer WSe$_2$ in sequence at a temperature of ~100 °C. Next, part of the graphene is picked up, and the stage is rotated to a desired twist angle. The remaining part of the graphene is then picked up, followed by h-BN and bottom graphite. The alignment angle between twisted bilayer graphene and monolayer WSe$_2$ is not controlled during the transfer process. Standard e-beam lithography is used to define the etch mask. CHF$_3$/O$_2$ and O$_2$ plasma etching are used to create a Hall bar geometry. Another round of e-beam lithography and e-beam evaporation is used to define electrodes (Cr/Au) that connect to the Hall bar. An optical image of the sample is shown in Fig. A1a. Throughout, the carrier density in the sample is controlled via external source meters (Keithly 2400).

The capacitance of each gate is confirmed via Hall measurements in a closed-cycle helium cryostat (Quantum Design) with a base temperature of 2 K. Standard lock-in techniques with a constant current (1-5 nA A.C., 13.3 Hz or 13.7 Hz) are employed for the measurements. The capacitance of each gate is determined by linearly fitting the gate-dependent Hall resistance ($R_{xy}$) using $R_{xy} = \frac{B}{\rho_H e}$ and $\frac{d\rho_H}{dV_g} = \frac{C_g}{e}$, where $B$ is the magnetic field, $\rho_H$ is the Hall carrier density, $V_g$ is the gate voltage, and $C_g$ is the gate capacitance. The twist angle is determined through the relationship $\rho = \frac{8\theta^2}{\sqrt{3}a^2}$ for integer fillings where the Hall effect onsets, assuming a graphene lattice constant of 0.246 nm. See Fig. A1b-e for corresponding transport data.

**Spectroscopic Measurements**

All spectroscopic measurements are performed in a closed-cycle liquid helium cryostat (Montana Instruments) under vacuum (<$10^{-6}$ torr). For the steady-state reflectance measurements, a 2796 K halogen lamp (SLS201L, Thorlabs) serves as the white light source. The lamp output is first cleaned up by coupling it to a single-mode fiber and collimating it with a triplet collimator. Then the output is focused onto the sample using a 40X, 0.6 NA objective. The beam waist is ~1 µm. The excitation power of the white light on the sample is estimated to be approximately 2 nW. The reflected light from the sample is next directed into a spectrometer and collected with a CCD array (PyLoN, Princeton Instruments). The steady-state reflectance contrast, $\Delta R/R$, is defined as $\Delta R/R = (R - R_0)/R_0$, where $R$ denotes the reflection spectrum on the sample and $R_0$ denotes the background reflection spectrum. $R$ is small for the moiré Rydberg states of WSe$_2$. To obtain $\Delta R/R$, the reference $R_0$ is taken at the same spot on the sample but with the sample gate tuned to a high doping level. This background reference is updated every 10 minutes to minimize background drift.

The transient reflectance experiments are seeded by a femtosecond laser (Carbide, Light Conversion) operating at 400 kHz and providing 230 fs pulses centered at 1.20 eV. A fraction of the output is then split to form the two paths for the pump and probe arms. For each path, the fundamental is attenuated and focused into a 4 mm yttrium aluminum garnet (YAG) crystal to generate a stable white light continuum. For the pump (probe), the continuum is spectrally filtered to 1.91~2.07 eV (1.71~1.82 eV) and further attenuated. We note that the photon energy of the pump is intentionally blue shifted with respect to that of the probe to avoid coherent artifacts otherwise arising in the fully collinear geometry when employing degenerate pulses. The pump is directed towards a motorized delay stage to control the time delay, $\Delta t$, and passed through an optical chopper ($f_{chopper}$ = 23 Hz) to generate alternating pump-on and -off signals. Both the pump and probe arms are directed collinearly to the sample through an objective (100X, 0.75 NA). The pump and probe spot diameters are ~1.1 µm and ~0.9 µm, respectively. Due temporal broadening from the objective, the instrument response function (IRF) is ~100 fs. The same objective is used to collect the reflected light which is spectrally filtered to remove the pump and dispersed onto a CCD detector array (PyLoN, Princeton Instruments). The pump-on and -off spectra at varying $\Delta t$ are then used to calculate the transient reflectance signal ($\Delta R/R$) where $\Delta R = R_{on} - R_{off}$ with the subscript indicates either pump-on or -off spectra.

**Data Analysis and Further Discussion**

In the following we describe the data processing used to analyze the overall transient signal and determine the $\rho$-dependent $k_{relaxation}$ and $k_{recovery}$ response. To aid visualization given the sparse $\rho$ sampling, the data used in Fig. 1d-g are first smoothed along the x-axis using a Savitzky-Golay filter, and then plotted as a function of probe energy versus $\rho$ at a fixed $\Delta t$.

To extract the peak positions, the data are first plotted as a function of probe energy versus $\Delta t$ at a fixed $\rho$ (see pseudocolor plots in Fig. 2 and 3). To carefully determine the $\Delta t$-dependent excitonic peak positions, these data are first smoothed along the y-axis using a Savitzky-Golay filter, producing the processed data shown in Fig. A2 and A3. The peak position is taken to be the probe energy corresponding to the minimum signal value (by convention, photobleach features are negative). The extracted peak positions are overlaid on the pseudocolor plots in Fig. 2 and 3. To account for uncertainty in the peak determination, error bars (also shown in Fig. 2 and 3) indicate the energy range that falls within 95% of the minimum signal. An example of this analysis for $\rho = -2.3 \times 10^{-12}$ cm$^{-2}$ at $\Delta t = 2.8$ ps is shown in Fig. A4. Data points for which the error range falls within 5 meV of the central peak are considered unreliable due to potential spurious experimental noise and are excluded from further analysis. This issue primarily arises under conditions of large electron doping, where the excitonic peak redshifts toward the edge of the probe window.

To generate Fig. 4a, the peak positions at each $\rho$ are fit using a linear function ($f = A \cdot k_{relaxation} + E_0$). The slope, $k_{relaxation}$, corresponds to the rate of energy relaxation (in units of meV/ps) of the peak position and is the parameter of interest. $E_0$ is the initial peak position at early $\Delta t$, with a negative slope indicating a dynamic redshift of the central peak position. Fig. A5 and A6 show the peak positions (as also overlaid on the pseudocolor plots in Fig. 2 and 3, respectively) under hole and electron doping along, along with the corresponding linear fits. For each carrier density, the fitted $k_{relaxation}$ value is shown (the associated error is reported as a 99% confidence interval). As a note, only data up to 7.5 ps were used in the fits for $\rho = -3.8 \times 10^{-12}$ cm$^{-2}$ (Fig. A5e), $-3.8 \times 10^{-12}$ cm$^{-2}$ (Fig. A5f), and $2.3 \times 10^{-12}$ cm$^{-2}$ (Fig. A6d) to avoid fit bias from data points near the edge of the probe window. These data points are still shown semi-transparently in Fig. A5 and A6.

Fig. A7 shows the peak amplitudes corresponding to each of the peak positions plotted in Fig. 2, 3, A5, and A6. Data under high electron doping ($\rho = 3.8 \times 10^{-12}$ cm$^{-2}$ and $4.5 \times 10^{-12}$ cm$^{-2}$) are excluded from this analysis due to insufficient data points at later Δt to reliably extract a recovery time constant. For the extracted data, error bars indicate signal range within 95% of the minimum signal (e.g., Fig. A4). To generate Fig. 4b, which presents the $\rho$-dependent $k_{recovery}$ (= $1/\tau_{recovery}$), these data were fit using the function $f = A \cdot \exp[t/\tau_{rise}] + B \cdot \exp[t/\tau_{relaxation}] + C$, convoluted with the IRF (~100 fs). The corresponding fits are overlaid on the data in Fig. A7 and the extracted $\tau_{recovery}$ values, with the associated 99% confidence intervals, are shown. Across the datasets, $\tau_{rise}$ falls within the range 0.5 ± 0.1 ps to 2 ± 0.2 ps. While various processes are known to occur on this timescale, as discussed in detail below, the observed dynamics likely arise from an overlapping photoinduced absorption feature related to unavoidable photoexcitation of tBLG.

In graphene, photoexcited carriers first thermalize <100 fs via carrier scattering [53,54]. Subsequently, this excess energy is dissipated from the electronic system to the lattice via optical phonons on a ~0.2 ps timescale and from optical to acoustic phonons on a ~2 ps timescale at which point interfacial thermal transport processes take effect [53,54]. As a note, thermal transport typically occur on much longer timescales (e.g., >10 ps for Gr to $WS_2$) than the observed transient redshifts in our transient measurements, and it has been shown to in fact be negligible for Gr to $WSe_2$ due to the poor heat transport properties of the latter [55] Furthermore, although such heating effects could certainly exhibit some $\rho$-dependence, a complete quenching of these processes at charge neutrality would not be expected. The rapid sub-ps thermalization timescale of the photoexcited carriers in tBLG also excludes possibility that the ps-scale dynamic peak evolution arises from photoinduced screening effects, i.e., the $WSe_2$ excitons experience a largely thermalized moiré charge distribution. While difficult to distinguish, interfacial charge transfer may also occur on a ~0.2 ps timescale, however, charge recombination follows shortly on a ~2 ps timescale [55–58].

**Appendix B: Extended Data**

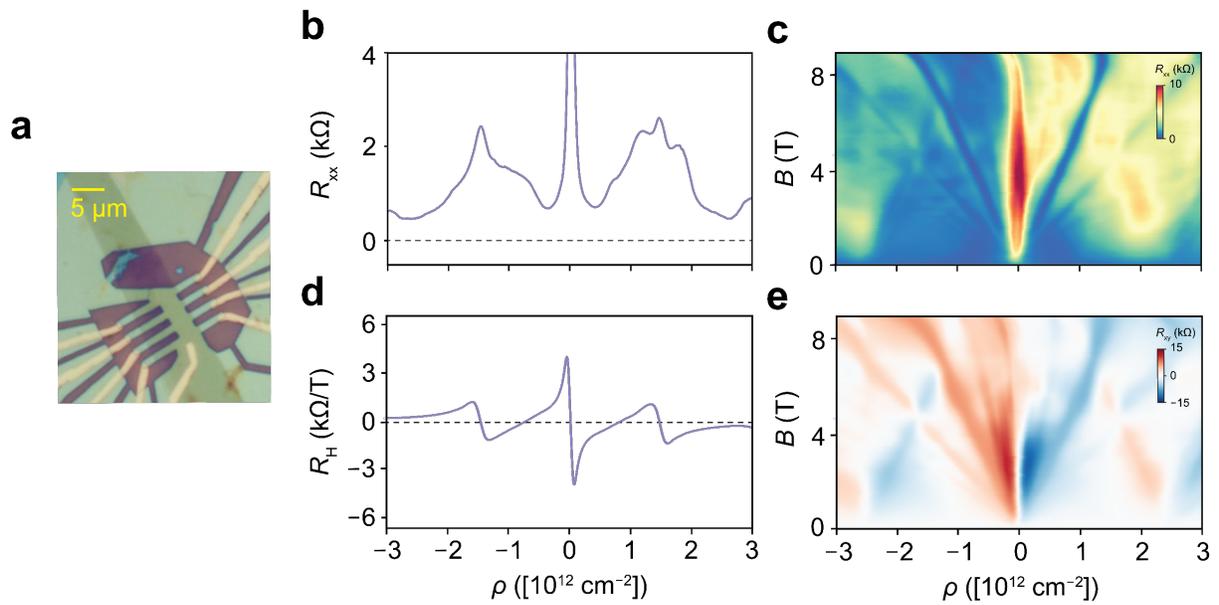

**Figure A1. Additional Device Characterization. a** Optical image of the device. **b-c** Resistance ($R_{xx}$) and Landau fan diagram of $R_{xx}$ as a function of carrier density ($\rho$), respectively. **d** Hall coefficient ($R_H$) as a function of $\rho$. **e** Landau fan diagram of the Hall resistance ($R_{xy}$). All data collected at 2 K.

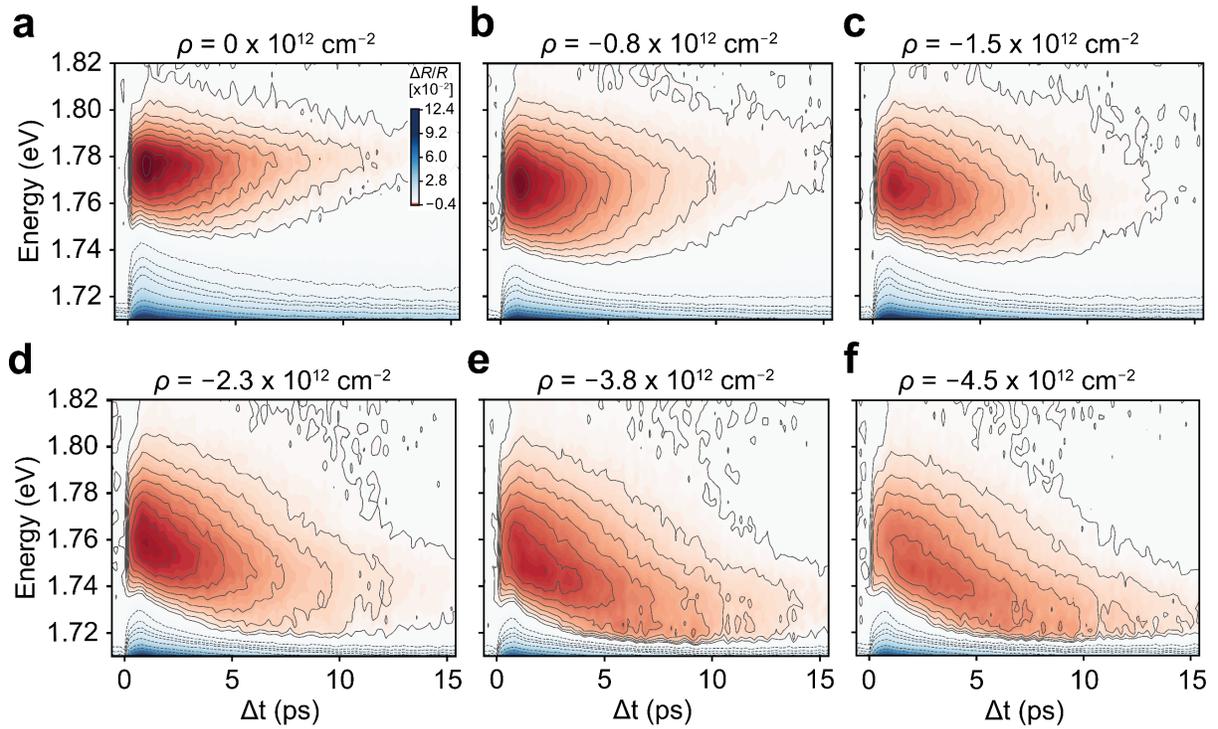

**Figure A2. Smoothed Transient Response of the Moiré Rydberg Exciton State as a Function of Hole Doping. a-f** Carrier density ($\rho$)-dependent exciton response shown as a function of probe energy versus pump-probe delay ($\Delta t$). As described in Methods, the data here have been smoothed along the y-axis with a Savitzky-Golay filter to facilitate peak finding. The corresponding unsmoothed data is shown in Fig. 2. Here, $\Delta R = R_{on} - R_{off}$ where the subscript indicates either pump-on or -off spectra. All data collected at T = 14 K.

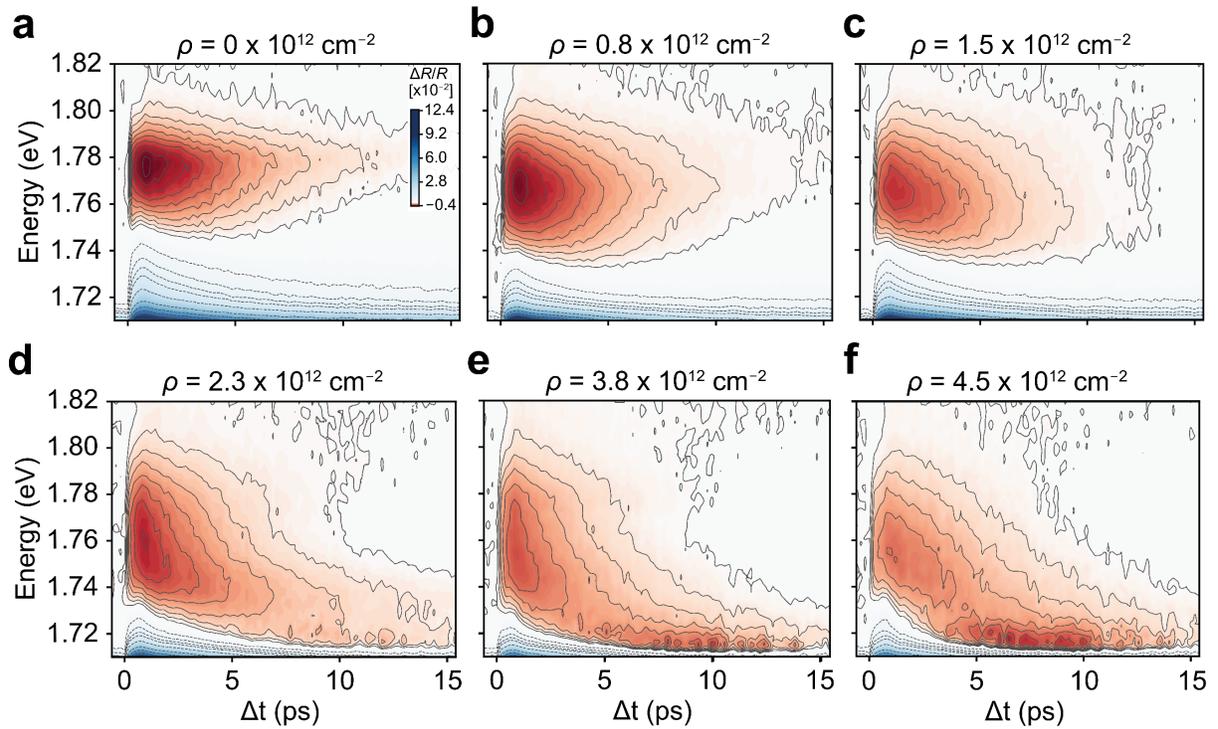

**Figure A3. Smoothed Transient Response of the Moiré Rydberg Exciton State as a Function of Electron Doping. a-f** Carrier density ($\rho$)-dependent exciton response shown as a function of probe energy versus pump-probe delay ($\Delta t$). As described in Methods, the data here have been smoothed along the y-axis with a Savitzky-Golay filter to facilitate peak finding. The corresponding unsmoothed data is shown in Fig. 3. Here, $\Delta R = R_{on} - R_{off}$ where the subscript indicates either pump-on or -off spectra. All data collected at T = 14 K.

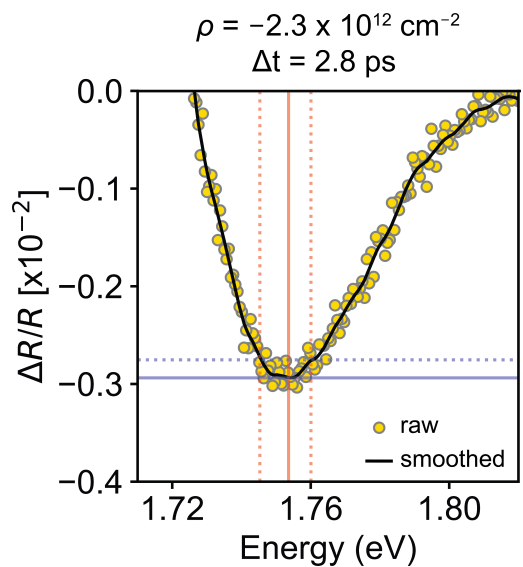

**Figure A4. Example of Peak Finding Analysis.** This panel shows the photobleach peak of the Rydberg exciton transition under $\rho$ = -2.3 x $10^{-12}$ $cm^{-2}$ at $\Delta t$ = 2.8 ps. Raw data points are shown as yellow circles and the smoothed data is shown by the solid black line. The peak minimum is identified at the energy marked by the vertical solid red line, with its corresponding amplitude indicated by the horizontal solid blue line. Errors are estimated by determining the energy and amplitude ranges that fall within 95% of the peak values (dotted lines).

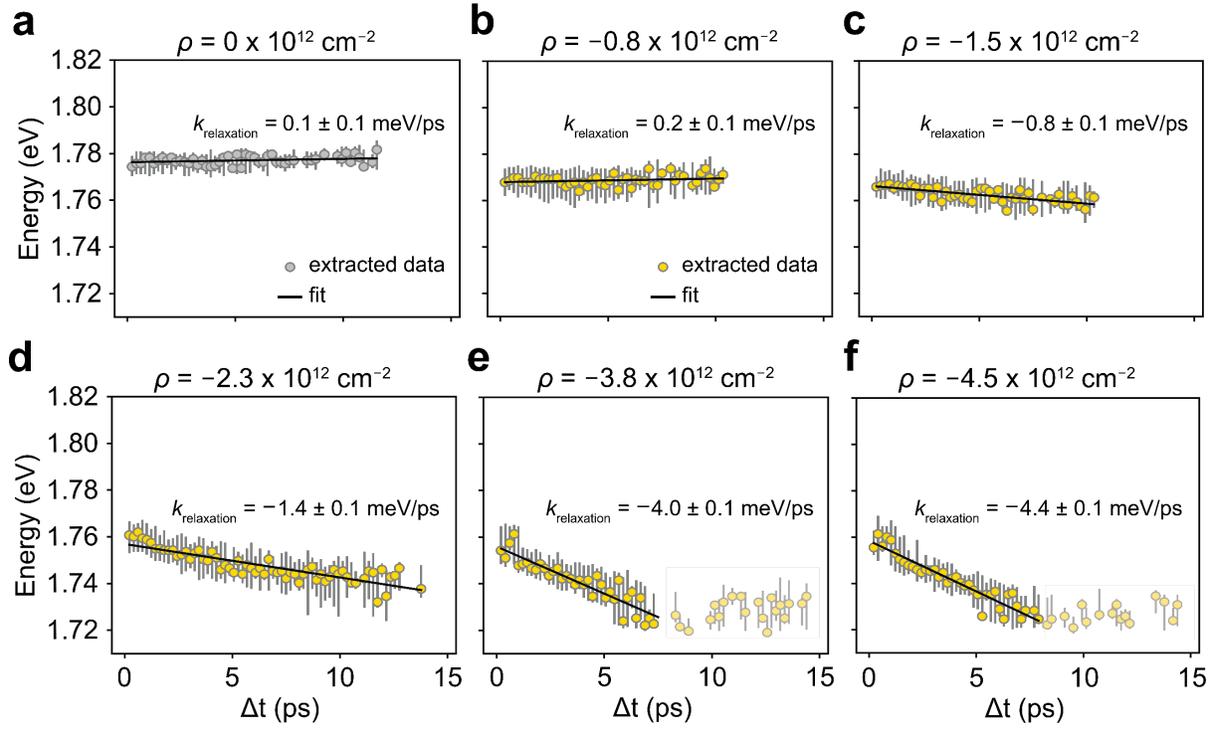

**Figure A5. Determination of $k_{relaxation}$ under Hole Doping. a-f** Carrier density ($\rho$)-dependent exciton peak shift as a function of pump-probe delay ($\Delta t$). The extracted peak positions at each $\Delta t$ are shown as gray circles (charge neutral) or yellow circles (hole doping). These data are also overlaid on the pseudocolor plots in Fig. 2. Error bars are calculated as described in Methods and illustrated in Fig. A4. The linear fits used to extract $k_{relaxation}$ are shown as solid black lines, with the corresponding fit values and 99% confidence intervals indicated in each panel. In panels e and f, the semi-transparent data has been excluded from the analysis as described in Methods.

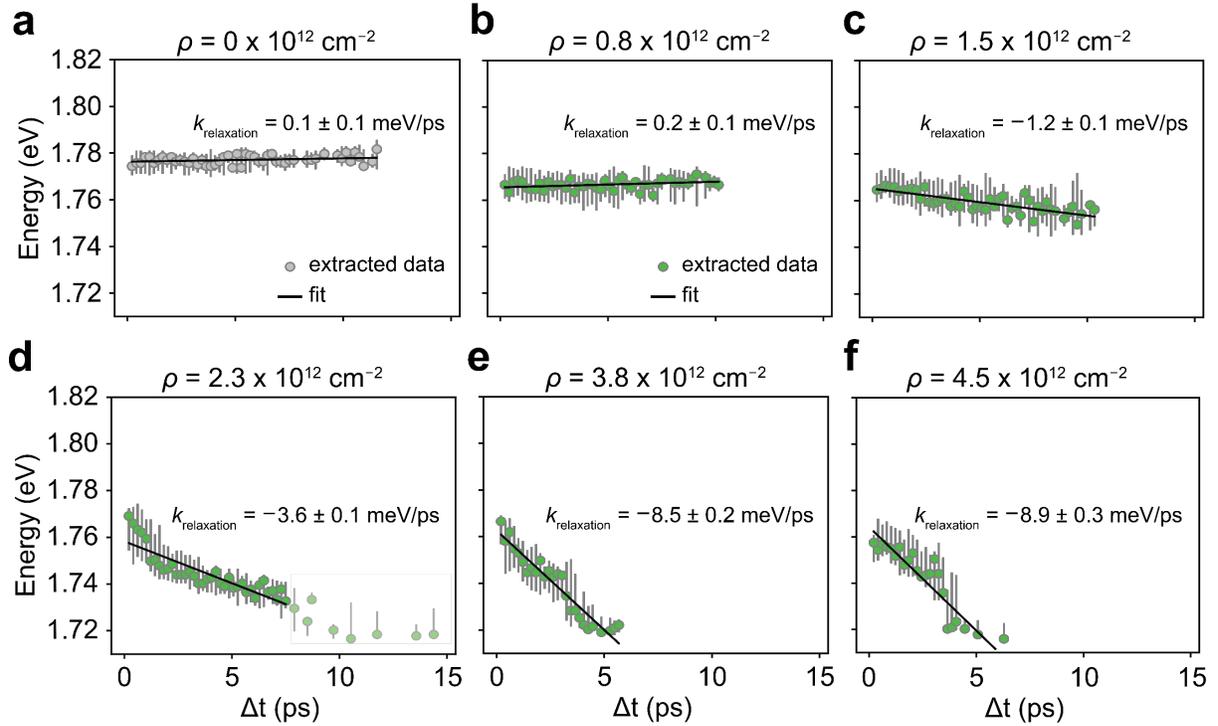

**Figure A6. Determination of $k_{\text{relaxation}}$ under Electron Doping. a-f** Carrier density ($\rho$)-dependent exciton peak shift as a function of pump-probe delay ($\Delta t$). The extracted peak positions at each $\Delta t$ are shown as gray circles (charge neutral) or green circles (electron doping). These data are also overlaid on the pseudocolor plots in Fig. 3. Error bars are calculated as described in Methods and illustrated in Fig. A4. The linear fits used to extract $k_{\text{relaxation}}$ are shown as solid black lines, with the corresponding fit values and 99% confidence intervals indicated in each panel. In panel d, the semi-transparent data has been excluded from the analysis as described in Methods. We also note that panel a here is identical to panel a in Fig. A5 and has been repeated to facilitate more straightforward observation of the $\rho$-dependent behavior.

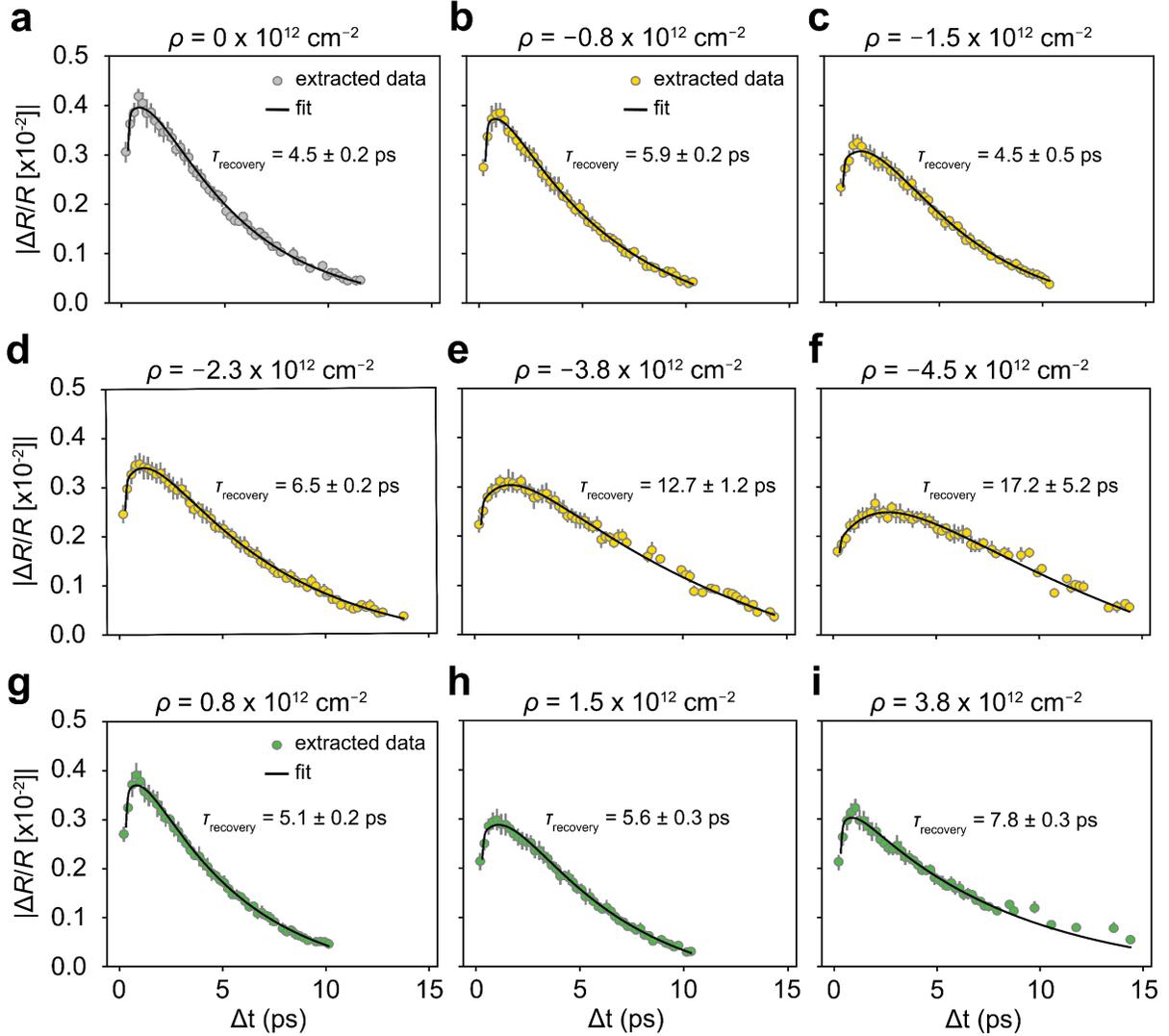

**Figure A7. Determination of $k_{recovery}$ (= $1/\tau_{recovery}$). a-i** Carrier density ($\rho$)-dependent exciton peak amplitude as a function of pump-probe delay ($\Delta t$). The extracted peak amplitudes at each $\Delta t$ are shown as gray circles (charge neutral), yellow circles (hole doping), or green circles (electron doping). These amplitudes correspond to the peak energies shown in Fig. 2, 3, A5, and A6. Error bars are calculated as described in Methods and illustrated in Fig. A4. The fits used to extract $\tau_{recovery}$ (=$1/k_{recovery}$) are shown as solid black lines, with the corresponding fit values and 99% confidence intervals indicated in each panel.